\documentstyle[12pt]{article}

\begin{document}
\begin{flushright}
SU-4240-532\\
UR-1306\\
ER-40685-755\\
March,1993
\end{flushright}
\def\cen{\centerline}
\vspace{25pt}
\cen{\Large {\bf HEAVY QUARK SOLITONS}}
\vspace{25pt}
\cen{\large Kumar S. Gupta}
\vspace{7pt}
{\small
\cen{ University of Rochester,}
\cen{ Department of Physics  and Astronomy ,}
\cen{ Rochester, NY 14627}}
\vspace{8pt}
\cen{ and}
\vspace{10pt}
\cen{\large M.Arshad Momen, J.Schechter and A. Subbaraman}
\vspace{8pt}
{\small\cen{ Department of Physics,}
\cen{ Syracuse University}
\cen{ Syracuse, NY 13244-1130}}
\normalsize
\vspace{40pt}
\cen{\Large \bf Abstract}
\vspace{30pt}
\noindent
   We investigate the heavy baryons which arise as solitonic excitations in a
``heavy meson" chiral Lagrangian which includes the light vector particles. It
is found that the effect of the light vectors may be substantial. We also
present a simple derivation which clearly shows the connection to the
Callan-Klebanov approach.

\newpage

\section{Introduction}
\setcounter{equation}{0}
    The problem of describing baryons containing heavy quarks
({\it c, b,} ...) as Skyrme type solitons may be considered as an extension of
the problem of describing strange ``light" baryons. An approximation
in which the strange quark is considered very heavy was actually
discussed some time ago by Callan and Klebanov$^1$ and others$^2$. In
this picture one imagines a stationary $K$-meson to which is bound
a nucleon treated as a soliton of the $SU(2)\times SU(2)$ effective
chiral Lagrangian. Of course it is not clear that the approximation
of treating the strange quark as ``heavy" is the best one; comparable
results may be obtained in other treatments$^{3,4,5}$ which consider all the
baryon octet members on the same footing. On the other hand, it
is reasonable to expect that the bound state approach is a
suitable one for the baryons containing $c$ and $b$ quarks.

    An application to the heavy quark baryons based on a fairly literal
extension of refs. $1-2$ has in fact been given$^6$. More recently it
has been realized that to go to the heavy quark limit one should do more
than just let the heavy mass get large. One should$^7$ at the same time keep
the heavy particle four velocity fixed and impose the Isgur-Wise
heavy spin symmetry. This has the consequence that a heavy vector
particle should always appear together in a multiplet with the
heavy pseudoscalar. The latter feature was not taken into account
in refs.$1-2$, 6 and so the problem was recently reexamined in  some papers
by a San Diego-Caltech group$^{8,9}$
 and by a Hanyang-Seoul-Jagellonian-Saclay-Stony Brook group$^{10,11}$.

    In the present note we shall consider the problem when the light
vectors are included in the chiral Lagrangian. It is known that
this makes both the mesonic and solitonic sector more realistic for
the light particles. It is also likely to be important for the
interaction of the heavy particles with the light ones; for example,
the semileptonic $D\rightarrow K^*$ transition appears to dominate
over $D\rightarrow K\pi$.

    In addition, we will give a method of calculation which appears a bit
simpler than the previous ones and which makes clearer the connection with
ref. 1. This may be of relevance since there is apparently some
disagreement between the two groups mentioned above.

\section{Effective Lagrangian}

\setcounter{equation}{0}

    The light part of the effective Lagrangian of pseudoscalars and
vectors, ${\cal L}_{light}$ which we shall employ has been discussed
in a number of places; see ref. 12 for details and ref. 13 for an
updating of the symmetry breaking part as well as further references.
The relevant light fields belong to the $3 \times 3$ matrix of
pseudoscalars, $\phi$ and to the $3 \times 3$ matrix of vectors,
$\rho_\mu$. It is convenient to define objects which transform simply
under the chiral group:
$$ \xi=exp(i\phi/F_\pi), \ \ \ \ \ \ U=\xi^2   $$
$$ A_\mu^L=\xi\rho_\mu\xi^{\dagger}+ {i \over\tilde
g}\xi{\partial}_\mu\xi^{\dagger},\ \ \
A_\mu^R=\xi^{\dagger}\rho_\mu\xi + {i\over{ \tilde g}}\xi^\dagger
\partial_\mu \xi,    $$
\begin{equation}\label{2.1}
F_{\mu\nu}(\rho)=\partial_\mu \rho_\nu -\partial_\nu \rho_\mu
 -i{\tilde g} \left[ \rho_\mu, \rho_\nu \right] ,
\end{equation}
where $F_\pi \approx 0.132$ GeV and $\tilde g \approx 3.93$ for a typical
fit. We can do without the explicit form of ${\cal L}_{light}$ here; it
is given in (2.6), (2.7) and (2.13) of ref.12 (wherein $g$ should be
replaced by $\tilde  g$). We also do not need the details of the $SU(3)$
symmetry breaking discussed in ref. 13. Actually we will restrict ourselves
to two flavors of light quarks for simplicity.

    The new feature associated with the interactions of the heavy
meson fields involves the use of a field$^{14}$ combining the heavy
pseudoscalar, $P^\prime$ and heavy vector, $Q_\mu^\prime$ at fixed
4-velocity, $V_\mu$ :
\begin{equation}\label {2.2}
H={{1-i\gamma_{\mu}V_\mu}\over 2}(i\gamma_5 P^\prime+i\gamma_\nu Q_\nu^\prime),
\ \ \  {\bar H}=\gamma_4 H^\dagger \gamma_4.
\end{equation}

Note that $H$ is here considered to have dimension of mass. $H$ is a $4 \times
4$ matrix in the Dirac space and it also carries an unwritten light
flavor index. The chiral interactions of $H$ with the light pseudoscalars
were discussed in ref. 15. The extension to including light vectors was given
in
 ref. 16, which notation we precisely follow here, and in ref. 17.
The total light-heavy interaction terms we shall require are$^{16}$:
$$ {{{\cal L}_{heavy}}\over {M}}=i V_\mu Tr\left
[H(\partial_\mu-i\alpha{\tilde g}\rho_\mu
-i(1-\alpha)v_\mu)\bar H \right ] +id Tr \left[H\gamma_\mu \gamma_5 p_\mu
\bar H \right ]    $$
\begin{equation}\label {2.3}
+  {ic \over m_v} Tr \left [H\gamma_\mu \gamma_ \nu
F_{\mu \nu}(\rho) \bar H \right ],
\end{equation}
\noindent
where $m_v \approx 0.77$ GeV is the light vector mass and

\begin{equation}\label{2.4}
v_\mu, p_\mu= {i \over 2} (\xi \partial_\mu \xi^{\dagger} \pm
\xi^{\dagger} \partial_\mu \xi).
\end{equation}

Furthermore $M$ is the heavy meson mass. $\alpha$, $c$ and $d$ are
dimensionless coupling constants for the light-heavy interactions. The
choice $\alpha=1$ corresponds to a natural notion of light vector meson
dominance, which is interesting to test. As seems appropriate for
an initial treatment, we do not include terms in (\ref{2.3}) which are
higher order in $1/M$, contain more derivatives of the light
fields or involve light flavor symmetry breaking.

    For treating the light baryon as a soliton which gets bound
to the heavy meson, we shall need information about the classical
soliton solutions of ${\cal L}_{light}$. First note the $2\times 2$
matrix decomposition of $\rho_\mu$:
\begin{equation}\label{2.5}
\rho_\mu={1 \over \sqrt 2}(\omega_\mu {\bf 1} +\tau^a \rho_\mu ^a).
\end{equation}
Then we have (see e.g. ref. 12) the classical ``profiles":

$$ U_c=exp(i{\bf \hat x} \cdot {\mbox{\boldmath $\tau$}} F(r)), \ \ \ \
\omega_{0 c}= \omega(r),
 $$
\begin{equation}\label{2.6}
\rho^a_{ic}={1 \over {\sqrt 2 {\tilde g} r}}\epsilon_{ika}
{\bf \hat x}_k G(r),
\end{equation}
and $\omega_{ic}=\rho^a_{0c}=0$. The boundary conditions for a finite
energy baryon number one solution are:
$$ F(0)=-\pi, \ \ \ G(0)=2, \ \ \ \omega^\prime(0)=0, $$
\begin{equation}\label{2.7}
F(\infty)=G(\infty)=\omega(\infty)=0.
\end{equation}
Typically, the opposite sign is taken for $F(0)$ as a kind of convention
but the sign choice above gives the correct sign for both axial
vector current and vector current matrix elements$^{18}$.

\section{Overview of the approach}
\setcounter{equation}{0}

    The first step in the bound state approach is to find the channel
in which the potential between the heavy meson and the nucleon as
soliton is maximally attractive. Of course in order to get a positive
parity {\it heavy} baryon we should consider the orbital angular
momentum  $l$ to be unity. In ref. 1, the attraction is found in the
state with ${ l}=1$ and grand spin, ${\bf G}={1\over 2}$. The grand spin
is the sum of the isotopic spin of the heavy meson, $\bf{I}$ and the  angular
momentum, $\bf{J}$; it enters in the first place because the Skyrme soliton
ansatz, given in (\ref{2.6}), mixes the isospin and ordinary spaces.
We would like to display a suitable wavefunction for the heavy meson,
$\bar {H}$ in the $l\,=\,1$, G = ${1\over2}$ channel.
First note from (\ref{2.2}) that when we
specialize the heavy meson to be {\it at rest} ($V_i=0$), the $4\times 4$
matrix $\bar{H}$ has non-vanishing elements only in the lower left
$2\times 2$ subblock. The first index of this submatrix represents the
spin of the light degrees of freedom within the heavy meson while the
second index represents the spin of the heavy quark. Furthermore there
is the unwritten bivalent isospin index. Specializing to this
subblock we then write the $\bar{H}$ wavefunction as:
\begin{equation}\label{3.1}
{\bar H}^b_{ l  h}=
({\bf \hat x} \cdot {\mbox{\boldmath $\tau$}})_{bd}\,{{\epsilon_{ld}}
\over {\sqrt 2}}{{u(r)}\over{\sqrt{4\pi M}}}
\end{equation}
Here, the radial wave function, $u(r)$ is taken in the first approximation
to be localized at the origin, $r^2 u^2(r) \approx \delta(r)$. Note that
the quantity ${\bf \hat x}$ represents the angular part of the space
wave function and the first factor couples it to the isospin index, $b$
to give G = ${1\over2}$. In turn this is coupled to the light spin index,
${ l}$ with the Clebsch-Gordan coefficient ${1\over {\sqrt 2}} \epsilon_{ld}$
 to give G = 0.
Finally, the heavy spin index, ${ h}$ is left uncoupled (as
appropriate to the heavy spin symmetry) to give the desired
net result G = ${1\over 2}$.

    To see that the wavefunction (\ref{3.1}) gives attraction we consider the
matrix element of the ``potential" obtained from (\ref{2.3}):

\begin{equation}\label{3.2}
V=-iMd \int d^3x Tr(H\gamma_i\gamma_5 p_i\bar H)+ \cdot \cdot \cdot,
\end{equation}
where only the $d$ term has been shown for simplicity. Substituting in
(\ref{3.1}), $p_i$ from (\ref{2.4}) and (\ref{2.6})
as well as $\gamma_i\gamma_5 \rightarrow -i\sigma_i$ yields

\begin{equation}\label{3.3}
V={{d}\over{2}} F^\prime (0) \epsilon_{sc}(\sigma_i)_{s{ l}}
\left\{ ({\bf \hat x} \cdot {\mbox{\boldmath $\tau$}})_{ca}
[-{1 \over 2} \tau_i +
\hat x_i {\bf \hat x} \cdot {\mbox{\boldmath $\tau$}}]_{ab}
( {\bf \hat x} \cdot {\mbox{\boldmath $\tau$}})_{bd}
\right \} \epsilon_{{l}d} +\cdot \cdot \cdot.
\end{equation}
Noticing that the object within the curly brackets is simply ${1 \over 2}
(\tau_i)_{cd}$ we easily get
\begin{equation}\label{3.4}
V=-{{3d}\over 2}F^\prime (0)+ \cdot \cdot \cdot.
\end{equation}
This provides an attractive force in the $ l=1$, G = ${1 \over 2}$ channel
since $F^\prime(0)>0$ and $d$ is also expected to be positive.
We will give the contributions to $V$ from the first and third
terms in (\ref{2.3}) in the next section; they do not appear to
change the attraction to repulsion.
However if we were to replace the Clebsch-Gordan coefficient in (\ref{3.1})
by $\delta_{{l}1} \delta_{d1}$, which corresponds to coupling everything
but the heavy quark spin to G=1, we would find that the  $-{3 \over 2}$
factor in (\ref{3.4}) is replaced by $+{1\over 2}$ and repulsion.
This is in agreement with refs. $8-9$; in our approach the
 ${\bf \hat x}\cdot{\mbox{\boldmath $\tau$}}$ factor
appears in a very natural way and also there is no need to introduce
a collective rotation factor at the present stage.

The next step is to introduce collective coordinates by allowing those
constant transformations which leave the Lagrangian invariant to depend
on time. First consider the soliton ``rotational" modes,$^{19}$ $A(t)$;
these are defined from the transformations
$$ U({\bf x},t) = A(t) U_c({\bf x}) A^{-1}(t); \> \>
{\mbox{\boldmath $\tau$}} \cdot {\mbox{\boldmath $\rho$}}_i({\bf x},t) =A(t)
{\mbox{\boldmath $\tau$}} \cdot {\mbox{\boldmath $\rho$}}_{ic}(x)A^{-1}(t);
$$
\begin{equation}\label{3.5}
{\bar H}({\bf x},t) = A(t) {\bar H}_{bound} \, ,
\end{equation}
where ${\bar H}_{bound}$ is effectively the wavefunction in (\ref{3.1}) and
$A(t)$ acts on its isospace index. Defining the angular velocity,
${\bf \Omega}$ by $ A^\dagger {\dot A} = {i\over 2} {\bf \Omega} \cdot
{\mbox{\boldmath $\tau$}}  $ and substituting  (3.5)  into
$\int d^3 x  ({\cal L}_{light} + {\cal L}_{heavy}  ) $
gives an additional contribution  to the Lagrangian of the general
form$^1$:
\begin{equation}\label{3.6}
\delta L = {1 \over 2} K \, {\bf \Omega}^2 - \chi \, {\bf \Omega} \cdot
{\bf G}^H \> ,
\end{equation}
in which $ K$ and $\chi$ represent spatial integrals over the profiles in
(\ref{2.6}) while ${\bf G}^H$, to the accuracy of interest, is the expectation
value of the heavy particle grand-spin, ${\bf G}^H = \langle
{\mbox{\boldmath $\tau$}} /2 \rangle + \cdot \cdot \cdot $ . Note that
at the collective Lagrangian level, as discussed in ref. 1, the angular
momentum of the heavy field is represented by its grand-spin while the
heavy field isospin is transmuted to zero. From (\ref{3.6}), with the soliton
angular momentum, $J_i^s = \partial \delta L / \partial \Omega_i$,
the rotational collective Hamiltonian is obtained as
\begin{equation}\label{3.7}
H_{coll} = {1 \over{2K}}\left ( {\bf J}^s + \chi \, {\bf G}^H \right )^2 \,.
\end{equation}
The moment of inertia, $K$ is identified from the light soliton sector
as $K = {3 \over 2}(m_\Delta - m_N)^{-1}$ in terms of the nucleon and
$\Delta$ masses. Eqn (\ref{3.7}) can be simplified by noting that the total
angular momentum is given as ${\bf J} = {\bf J}^s + {\bf G}^H$. Then we
arrive at the heavy baryon mass formula$^1$:
\begin{equation}\label{3.8}
H_{coll} = {1 \over 3}(m_\Delta - m_N)\left [ (1-\chi)I(I+1) +
\chi \, J(J+1) + \cdot \cdot \cdot \right ] \>,
\end{equation}
where the three dots stand for a term which does not split the heavy
baryon masses. In deriving (\ref{3.8}), use was made of
the fact that the soliton
carries the full isospin of the heavy baryon at the collective level
so that ${\bf J}^2 = {\bf I}^2$. We will see that ${\cal L}_{heavy}$ in
(\ref{2.3}) leads to $\chi = 0$. Thus we have the final results:
\begin{equation}\label{3.9}
m(\Sigma^*_Q) - m(\Sigma_Q) = 0; \> m(\Sigma_Q) -m(\Lambda_Q)= {2 \over 3}
(m_\Delta - M_N) \, ,
\end{equation}
wherein the subscript $Q$ denotes the heavy baryon in which the $s$ quark
has been replaced by the heavy quark $Q$. The equality $m(\Sigma^*_Q) =
m(\Sigma_Q)$ is, of course, expected from the heavy quark spin symmetry.

Finally, we take into account the translational mode, ${\bf R}(t)$ for
the soliton by now setting
\begin{equation}\label{3.10}
U({\bf x},t) = A(t) U_c\left ({\bf x}-{\bf R}(t)\right ) A^{-1}(t) \, ,
\end{equation}
and similarly treating the light vector fields.  Imagining the heavy meson
at rest with the soliton moving about it, we have a Schr\"odinger
equation for the relative motion
\begin{equation}\label{3.11}
\left ( - {1 \over {2m_N}} \nabla_R^2 + V({\bf R}) \right )u({\bf R}) =
E\, u({\bf R}) \, ,
\end{equation}
wherein $V({\bf R})$ is to be identified with (\ref{3.4}).
It will be expanded to quadratic order
\begin{equation}\label{3.12}
V({\bf R}) = V_0 + {1 \over 2} \kappa \, {\bf R}^2 \cdot \cdot \cdot \, ,
\end{equation}
following the notation of ref.  9.

\section{Effects of light vector fields}
\setcounter{equation}{0}
    Here we give details of the calculation, including all three
terms in ${\cal L}_{heavy}$, (\ref{2.3}). The $\alpha$ term involves the
profile $\omega (r)$ and the $c$ term involves the profile $G(r)$.
The needed factors for the non zero contributions to the
``potential", $V(r)$ are
$$ \rho_0 = {1 \over {\sqrt 2}} \omega (r), \ \ \ p_i =
{{\sin F}\over{2r}} (\tau_i -\hat x_i \, {\bf \hat x} \cdot
{\mbox{\boldmath $\tau$}} )
+{1 \over 2}F^\prime \hat x_i\,{\bf \hat x} \cdot {\mbox{\boldmath $\tau$}},$$
\begin{equation}\label{4.1}
\gamma_i \gamma_j F_{ij}(\rho) = {i \over {{\tilde g} r}} \left[
-G^\prime {\mbox{\boldmath $\sigma \cdot \tau$}} +{1 \over r} ( r G^\prime +
G(G-2)) {\bf \hat x} \cdot {\mbox{\boldmath $\sigma$}} \,
{\bf \hat x} \cdot {\mbox{\boldmath $\tau$}} \right].
\end{equation}
We expand the profiles around the origin as follows:
$$F(r)=-\pi + r F^\prime(0) + {1 \over 6}r^3 F^{\prime \prime \prime}(0)
+ \cdot \cdot \cdot, $$
$$ \omega(r) = \omega (0) + {1 \over 2 } r^2 \omega^{\prime \prime }(0)
+ \cdot \cdot \cdot , $$
\begin{equation}\label{4.2}
G(r)= 2 + {1 \over 2} r^2 G^{\prime \prime}(0) +
{1 \over {24}}r^4 G^{\prime \prime \prime \prime}(0) + \cdot \cdot \cdot,
\end{equation}
where powers were deleted so that the underlying fields are analytic$^9$
in the Cartesian coordinates. Substituting (\ref{4.1}) and (\ref{4.2}) into
(\ref{2.3}) and evaluating the resulting expression in the attractive channel
corresponding to wavefunction (\ref{3.1}) enables us to identify
the parameters of the ``potential" as
\begin{equation}\label{4.3}
V_0 = - { 3 \over 2} dF^\prime(0) +{ {3c} \over {m_v {\tilde g}}} G^
{\prime \prime} (0) -{{\alpha {\tilde g}} \over {\sqrt 2}} \omega (0),
\end{equation}
\begin{equation}\label{4.4}
\kappa = d (-{5 \over 6} F^{\prime \prime \prime}(0) +{1 \over 3}
F^\prime (0)^3) + {c \over {m_v {\tilde g}}} ( {5 \over 6}
G^{\prime \prime \prime \prime}(0) + {1 \over 2} G^{\prime \prime}
(0)^2 ) -{{ \alpha {\tilde g}} \over {\sqrt 2}} \omega^{\prime \prime}(0) .
\end{equation}
If we choose to evaluate $V(r)$ in the ``repulsive" channel
mentioned after (\ref{3.4}), the $c$ and $d$ terms in both (\ref{4.3})
and (\ref{4.4}) would be multiplied by $-1/3$ while the $\alpha$ terms
would remain unchanged.

    It is of course necessary to know the derivatives of the $F(r)$,
$G(r)$ and $\omega(r)$ profiles which appear above. These are  found by
solving the coupled differential equations which arise from
minimizing the static energy of ${\cal L}_{light}$ (see (6.6)
of ref 12) subject to the boundary conditions (\ref{2.7}) above. For
the typical best fit to meson and light baryon masses these are
uniquely characterized by the starting values$^{20}$:
\begin{equation}\label{4.5}
F^{\prime}(0)=0.795 \, GeV, \ \ \ G^{\prime \prime}(0)=-0.390 \, GeV^2,
\ \ \ \omega(0)=-0.094 \, GeV.
\end{equation}
{}From (\ref{4.5}) and the differential equations it is straightforward
to deduce the next higher derivatives:
\begin{equation}\label{4.6}
F^{\prime \prime \prime}(0)=-0.12 \, GeV^3, \ \ \ G^{\prime \prime
\prime \prime}(0)=0.35 \, GeV^4, \ \ \ \omega^{\prime \prime}(0)=
0.016 \, GeV^3.
\end{equation}

    Finally, let us justify the vanishing of the $\chi \, {\bf \Omega}
\cdot \langle {\mbox{\boldmath $\tau$}}\rangle /2$ term in (\ref{3.6})
which was used to obtain
the mass formulas in (\ref{3.9}). This term might arise when (\ref{3.5})
is substituted
into (\ref{2.3}). We note that a possible contribution from the $d$ term
would involve a factor $p_4$ and hence a factor $ \gamma_4 \gamma_5$,
which has no components in the $2 \times 2$ subblock for which ${\bar H}
H$ is non-vanishing. Similarly the relevant factor for the $c$ term
involves a factor $\gamma_k \gamma_4$, which also has a vanishing overlap
with ${\bar H}H$. For the first term in (\ref{2.3}) we may
explicitly check that the operator {\boldmath $\tau$} has zero
expectation value in the wavefunction (\ref{3.1}).

\section{Results and discussion}
\setcounter{equation}{0}
    To gauge the accuracy of the heavy quark symmetry we may observe that
the vector-pseudoscalar mass differences,which should vanish in the
$M \rightarrow \infty$ limit, are respectively about 400 MeV, 145
MeV, and 45 MeV for $K^*-K, D^*-D$ and $B^*-B$. From this point of view,
the s-quark should not be considered heavy while it seems quite reasonable
to treat the $b$-quark in this manner.

    First let us look at the mass splitting relations (3.9), which are
independent of the fine details of the model. Unfortunately there
is presently not enough heavy baryon data to check the formula $m(\Sigma^*_Q)-
m(\Sigma_Q)=0$; it is not surprising that one gets 193 MeV instead
of 0 when one uses the hyperons formed with the $s$-quark. For the
relation $m(\Sigma_Q) - m(\Lambda_Q) \approx 195 MeV$, one finds
77 MeV with the ordinary hyperons but 170 MeV with the $c$-quark
hyperons. The fair agreement in the latter case is encouraging.
The results in the former case show that the Callan-Klebanov model
should not be applied to the strange hyperons if one goes to the true
heavy meson limit.

    Additional physical quantities may be estimated with the help
of the ``potential" for the bound state problem; substituting (4.5)
and (4.6) into (4.3) and (4.4) yields:
$$V_0=-1.19d -0.39c +0.26\alpha \ \ GeV \, ,$$
\begin{equation}\label{5.1}
\kappa=0.27d +0.12c -0.04\alpha \ \ GeV^3,
\end{equation}
for the typical choice of light particle Lagrangian parameters.
This, of course, involves the three light-heavy coupling coefficients
$d$, $c$ and $\alpha$. The value of $d$ has been discussed in the
literature$^{21}$  ; bounds are obtained which agree with a
simple estimate$^{16}$ based on pole dominance of the $D \rightarrow K$
 transition form factor:
\begin{equation}\label{5.2}
d =\beta{{F_{\pi}}\over{F_D}} \approx 0.53,
\end{equation}
where $\beta \approx 1.0$ and we have used$^{22}$
$F_D \approx 250 MeV$. We shall adopt (5.2) for definiteness. In
contrast we are just at the initial stage for determining the
values of $c$ and $\alpha$; the present model may assist the
direct approach from meson weak transition matrix elements in
this regard.

    At lowest order, $V_0$  is  a measure of the energy
which binds the heavy meson and soliton together as a heavy
baryon. Allowing for a zero point energy $Z$  for quantum fluctuations,
 we write
\begin{equation}\label{5.3}
m(\Lambda_Q)-m_N -M = V_0 + Z .
\end{equation}
The left hand side is about $ -780\, MeV$ if we identify $Q=b$. If we
first neglect $Z$ and set $\alpha=1$ (``vector meson dominance"), we
can get an idea of the value of c by putting (5.1) into (5.3). This
gives us $c \approx 1.0$, which seems quite reasonable in that it
is expected to be of the same order of magnitude as $d$. Examining
the individual terms in (5.1) we see that the pion ($d$)
term is dominant but the $\rho$ meson($c$) contribution is 60\% as large
and also attractive while the omega ($\alpha$) contribution is
40\% as large and repulsive. Of course there can be different fits
but this seems to be the typical situation. It is clear that the
vector mesons can be expected to play a non-trivial role.

    Using the numbers given above we find the parameter $\kappa$,
which measures the strength of the $R^2$ part of the collective
mode potential in (3.12), to be about 0.22 $GeV^3$. With (3.12)
we note that (3.11) is just the Schr\"odinger equation for a spherical
harmonic oscillator with angular frequency $\left({{\kappa}\over {m_N}}\right)
^{1 \over 2}
\break \approx 0.5\, GeV$. As is well known this results in a zero point
energy,$Z \approx 0.75\, GeV$. Such a value is not a small perturbation
to the right hand side of (5.3) so one might expect the collective
ansatz in (3.10) to be suspicious. It would be more reliable to
include the full non-linearities of the soliton profiles as well as
to include higher derivative interactions in the light-heavy Lagrangian.
This is a future job, but we can at least mention
the parameters which will fit (5.3) to experiment while retaining
$Z={3 \over 2 }\left({{\kappa}\over {m_N}}\right)^{1 \over 2}$.
 Keeping $\alpha =1$ and $d=0.53$ would
result in $c=4.1$ and correspondingly $\kappa =0.6\, GeV^3$. Such a value of
$c$ is on the large side. An alternative fit is obtained by choosing
$c=1$ and taking $d=0.7$ (the experimental bound); then we find
$\alpha = -2$ (which does not agree with our intuitive notion of light vector
 meson dominance) and correspondingly $ \kappa =0.39$. The two kinds
of fits give different values for $\kappa$. Since the orbital
excitations of the spherical harmonic oscillator are equally spaced
at intervals given by the angular frequency, we would expect the  first
orbitally excited $ \Lambda_b$ to be about $0.80\, GeV$ higher for
the first fit while about $0.63\, GeV$ higher for the second fit.
Another possible future source of information about $\kappa$ is
the Isgur-Wise function for the $\Lambda_b \rightarrow \Lambda_c$
weak transition form factor. It was shown in
ref. 9 that the non-perturbative part of this form factor is
simply given by the overlap of the initial and final state
spherical harmonic oscillator wave functions:
\begin{equation}\label{5.4}
\eta_0(V \cdot V^\prime)= exp \left[({m_N}^3/ \kappa)^{1/2}
(V \cdot V^\prime +1) \right],
\end{equation}
where the strict $M \rightarrow \infty$ limit was taken for
simplicity of writing.

\vspace{5pt}

    We would like to thank Herbert Weigel for helpful
discussions. This work was supported in part by the U.S.
Department of Energy under the contract number DE-FG-02-85ER40231
and also by the grant number DE-FG02-91ER40685.

\vskip 25pt

\leftline{ {\Large{\bf References}} }

\vspace{20pt}

\begin{enumerate}
\item C.G.Callan and I. Klebanov, Nucl. Phys. B{\bf 262},365(1985);
C.G. Callan, K. Hornbostel and I. Klebanov, Phys. Lett. B{\bf 202}
(1988).

\item J. Blaizot, M. Rho and N. Scoccola, Phys. Lett B {\bf 209}
,27(1988); N. Scoccola, H. Nadeau, M. Nowak and M. Rho, {\it
{ibid.}}{\bf 201}, 425(1988);D. Kaplan and I. Klebanov, Nucl.
Phys. B{\bf 335},45(1990).

\item H.Yabu and K. Ando, Nucl. Phys. B{\bf 301}, 601(1988).

\item H. Weigel, J. Schechter, N. W. Park and Ulf-G. Meissner,
Phys. Rev. D{\bf 42}, 3177(1990).

\item N. W. Park and H. Weigel, Nucl. Phys. A{\bf 541},453(1992).

\item M. Rho, D. O. Riska and N. Scoccola, Z. Phys. A{\bf 341},
343(1992).

\item E. Eichten and F. Feinberg,Phys. Rev. D {\bf 23},2724(1981);
M. B. Voloshin and M. A. Shifman , Sov. J. Nucl. Phys. {\bf 45},292,(1987);
N. Isgur and M. B. Wise, Phys. Lett. B{\bf 232}, 113 (1989);
H. Georgi , Phys. Lett. B {\bf240}, 447 (1990).

\item E. Jenkins, A. Manohar and M. B. Wise,Caltech preprint CALT-68-1783-REV
(hep-ph/9205243);Z. Guralnik, M. Luke and A. V. Manohar,
Nucl. Phys. B{\bf 390},474(1993);
E. Jenkins and A. Manohar,Phys. Lett. B{\bf 294},173(1992).

\item E. Jenkins, A. Manohar and M. Wise, UCSD preprint UCSD/PTH 92-27
(hep-ph/9208248).

\item M. Rho, Talk at Workshop on ``Baryons as Skyrme Solitons",
Sept.28-30, 1992; Siegen, Germany (hep-ph /9210268).

\item D.P. Min, Y. Oh, B. Y. Park and M. Rho, Seoul preprint SNUTP 92-78
(hep-ph/9209275);
H. K. Lee, M. A. Nowak, M. Rho and I. Zahed, Stony Brook
preprint NTG-92-20( hep-ph/9301242).

\item P. Jain, R. Johnson, Ulf-G. Meissner, N. W. Park and J. Schechter,
Phys. Rev. D{\bf 37},3252,(1988).

\item J. Schechter, A. Subbaraman and H. Weigel, Syracuse-T\"ubingen
preprint SU-4240-525( hep-ph/9211239).

\item J.D. Bjorken, {\it in}  La Thuile Rencontres , 1990 , p.583;
M.Wise, Phys. Rev.D{\bf 45}, R2188(1992); F. Hussain , J.  K\"orner and
G. Thompson, Ann. Phys. {\bf 206},334,(1991) .

\item  M. Wise, ref. 14 above; T. M. Yan, H.Y. Cheng, C. Y. Cheung, G. L.
Lin, Y. C. Lin and H. L. Yu,Phys. Rev. D{\bf 46}, 1148(1992); G. Burdman
and J. F. Donoghue, Phys. Lett. B {\bf 280}, 280 (1992);
P. Cho, Phys. Lett. B {\bf 285}, 145 (1992).

\item J. Schechter and A. Subbaraman, Syracuse preprint SU-4240-519
(hep-ph/ 9209256).

\item R. Casalbuoni, A. Deandrea, N. Di Bartolomeo ,R. Gatto,
F. Feruglio and G. Nardulli , Phys. Lett. B{\bf 292}, 371(1992);
Phys. Lett. B{\bf 294}, 371(1992);Phys. Lett. B{\bf 299},139(1993);
N. Kitazawa and T. Kurimoto, Osaka preprint OS-GE-27-92 (hep-ph/9302296).

\item See Appendix B of N. W. Park, J. Schechter and H. Weigel,
Phys. Rev. D{\bf 41}, 2836(1990) and also Z. Guralnik et al,
ref. 8 above.

\item G. Adkins, C. Nappi and E. Witten, Nucl. Phys. B{\bf 228},
552(1983).

\item We are grateful to Herbert Weigel for supplying these. The
numbers given correspond to the choice of parameters $\gamma_2=
\gamma_3, |{\tilde g}_{VV\phi}|=1.9, |{\tilde h}|=0.4$, as discussed
in ref. 5 above, where slightly different conventions from the present
one were employed. Symmetry breaking terms were also included although
their effect on the profiles is small.

\item  P. Cho and H. Georgi, Phys. Lett. B{\bf 296},408(1992);
J. Amundsen et. al., {\it ibid}, 415(1992).

\item J. Amundsen, J. Rosner, M. Kelly, N. Horwitz and S. Stone,
Enrico Fermi Institute preprint, EFI 92-31 (hep-ph/9207235).

\end{enumerate}

\end{document}